\documentclass[journal=prm,manuscript=article,superscriptaddress,twocolumn]{revtex4-2}

\usepackage{color}
\usepackage{graphicx}
\usepackage[export]{adjustbox}
\usepackage[utf8]{inputenc}
\usepackage{todonotes}
\usepackage{amssymb}
\usepackage{amsmath}
\usepackage{physics}
\usepackage{float}
\usepackage{subfloat}
\usepackage{subfig}
\usepackage{gensymb}
\usepackage{mathtools}
\usepackage{ulem}
\begin{document}
\title{Excitation protocols for non-linear phononics in bismuth and antimony}
\author{Anubhab Haldar}
\affiliation{Department of Electrical and Computer Engineering, Boston University, Boston, MA 02215}
\author{Zhengjie Huang}
\affiliation{Center for Nanoscale Materials, Argonne National Laboratory, Argonne, IL 60439, USA}
\author{Xuedan Ma}
\affiliation{Center for Nanoscale Materials, Argonne National Laboratory, Argonne, IL 60439, USA}
\affiliation{Northwestern Argonne Institute of Science and Engineering, Evanston, IL 60208, USA}
\affiliation{Consortium for Advanced Science and Engineering, The University of Chicago, Chicago, Illinois 60637, United States}
\author{Pierre Darancet}
\email{pdarancet@anl.gov}
\affiliation{Center for Nanoscale Materials, Argonne National Laboratory, Argonne, IL 60439, USA}
\affiliation{Northwestern Argonne Institute of Science and Engineering, Evanston, IL 60208, USA}
\author{Sahar Sharifzadeh}
\email{ssharifz@bu.edu}
\affiliation{Department of Electrical and Computer Engineering, Boston University, Boston, MA 02215}
\affiliation{Division of Materials Science and Engineering, Boston University, Boston, MA 02215}
\date{\today}

\begin{abstract}
     We study the optical generation and control of coherent phonons in elemental bismuth (Bi) and antimony (Sb) using a classical equation of motion  informed by first-principles calculations of the potential energy surface and the frequency-dependent macroscopic dielectric function along the zone-centered optical phonons coordinates. Using this approach, we demonstrate that phonons with the largest optomechanical couplings, also have the strongest degree of anharmonicity among the zone-centered modes, a result of the broken symmetry structural ground state of Bi and Sb.  We show how this anharmonicity, explaining the light-induced phonon softening observed in experiments, prevents the application of standard phonon-amplification and annihilation protocols. We introduce a simple linearization protocol that extends the use of such protocols to the case of anharmonic phonons in broken symmetry materials, and demonstrate its efficiency at high displacement amplitudes. Our formalism and results provide a path for improving optical control in non-linear phononics.
\end{abstract}

\maketitle

Coherent interactions between electromagnetic waves and extended vibrational degrees of freedom in solids (phonons) enable the stabilization of highly non-thermal states of matter with potentially desirable properties \cite{delatorreRMPColloquiumNonthermalPathways2021}. Such interactions have been observed using femtosecond laser pulses in a wide variety of materials over the past 50 years, including oxides~\cite{chengAPLModulationSemicondutorSemimetal1993}, transition metal dichalcogenides~\cite{anikinPRBUltrafastDynamicsHighsymmetry2020}, elemental pnictogens~\cite{sharpJPFMPLatticeDynamicsAntimony1971,chengAPLImpulsiveExcitationCoherent1990,chengAPLMechanismDisplaciveExcitation1991,haseAPLOpticalControlCoherent1996,garrettPRLCoherentTHzPhonons1996,hasePRBDynamicsCoherentPhonons1998,decampPRBDynamicsCoherentControl2001,hasePRLDynamicsCoherentAnharmonic2002}, and have been leveraged for control of vibrational dynamics~\cite{dekorskyELCoherentControlLOPhonon1993,decampPRBDynamicsCoherentControl2001,haseAPLOpticalControlCoherent1996}, non-equilibrium structural phase transitions~\cite{chengAPLModulationSemicondutorSemimetal1993,anikinPRBUltrafastDynamicsHighsymmetry2020}, and to modulate the non-linear susceptibility~\cite{taghinejadPRLTransientSecondOrderNonlinear2020,haldarGiantOptomechanicalCoupling2021,khalsaPRXUltrafastControlMaterial2021}. 
%For these reasons, this optical control of the structural dynamics of materials is sometimes referred to as "non-linear phononics".

Microscopically, the coherent coupling between electromagnetic waves and vibrational degrees of freedom is generally understood as a consequence of vibration-induced changes in the dielectric susceptibility in the frequency range of the pump laser ~\cite{merlinSSCGeneratingCoherentTHz1997}. Such changes can be the result of processes involving Raman scattering~\cite{weberRamanScatteringMaterials2000}, infrared phonons \cite{dekorsyPRLEmissionSubmillimeterElectromagnetic1995, subediPRBTheoryNonlinearPhononics2014}, ionic Raman scattering \cite{subediPRBTheoryNonlinearPhononics2014,haldarGiantOptomechanicalCoupling2021}, or infrared resonant Raman scattering \cite{khalsaPRXUltrafastControlMaterial2021}. For opaque materials, an early phenomenological theory was the displacive excitation of coherent phonons (DECP)~\cite{zeigerPRBTheoryDisplaciveExcitation1992}, later shown to be a special case of impulsive stimulated Raman scattering~\cite{garrettPRLCoherentTHzPhonons1996}. The DECP model can explain the oscillatory cosine-like dependence of optical reflectivity upon excitation by impulsive laser sources by considering the change in the electron density in the presence of a coherently excited phonon. Such conclusions were also verified using a density-matrix-based model~\cite{stevensPRBCoherentPhononGeneration2002}  in the case of antimony. In contrast to this deep theoretical understanding of the underlying coupling mechanism, however, most illumination protocols leading to the stabilization of non-equilibrium phases have been discovered through trial-and-error approaches. A primary reason is the lack of quantitative methods capable of treating the full, anharmonic, potential energy surface and the driving force on an equal footing. To this end, first-principles approaches using Born-Oppenheimer~\cite{khalsaPRXUltrafastControlMaterial2021} or Ehrenfest~\cite{lloyd-hughesJPCM2021UltrafastSpectroscopic2021,xuSADecoupledUltrafastElectronic2022} dynamics in conjunction with time-dependent density functional theory (TDDFT) provide an important opportunity in the understanding and control of these systems.

In this work, using a fully first-principles-informed classical model, we study light-induced structural dynamics in bismuth (Bi) and antimony (Sb) --two broken-symmetry elemental solids, which have been extensively studied experimentally and show clear macroscopic evidence of coherent phonons \cite{merlinSSCGeneratingCoherentTHz1997, hasePRBDynamicsCoherentPhonons1998, hasePRLDynamicsCoherentAnharmonic2002,murrayPRBEffectLatticeAnharmonicity2005,murrayPRBPhononDispersionRelations2007,papalazarouPRLCoherentPhononCoupling2012,misochkoJETPExperimentalEvidenceExistence2014}. For these systems, we calculate the potential energy surface (PES) and the frequency-dependent dielectric function within density functional theory (DFT) and TDDFT, respectively, along the full range of phonon amplitudes, $Q_i$, for the zone-centered optical modes. With these parameters as input, we perform classical Born-Oppenheimer dynamics simulation with a non-linear force term explicitly depending on phonon coordinates and electric field magnitude.  Using this approach, we demonstrate that the $A_{1g}$ phonon mode -- the mode with the largest optomechanical couplings in Bi and Sb, also has the strongest degree of anharmonicity among the zone-centered modes, explaining the light-induced phonon softening observed in experiments. Such anharmonicity results from the broken-symmetry structural ground state and prevents the application of standard phonon-amplification and annihilation protocols based on a period/half-period pulse train. We introduce a simple linearization protocol that extends the use of phonon-annihilation and amplification protocols to the case of such phonons, and demonstrate its efficiency at high displacement amplitudes. 
%Our formalism and results provide a path for improving optical control in non-linear phononics in broken symmetry materials.

%\section{Computational and Experimental Details}
%\section{Computational Details}
%All numerical details, code, and input files are provided with the Supplementary information. 
Briefly, the set of phonon coordinates $Q_i(t)$ (where $i$ is the mode index) at an instant $t$ are evolved on the Born-Oppenheimer energy surface, using the following instantaneous forces, $\mathbf F(\mathbf Q_i , t)$,
\begin{widetext}
\begin{align}
    \mathbf F(\mathbf Q_i ,  t)  = - \left.\frac{dU}{d \mathbf Q}\right|_{\mathbf Q=\mathbf Q_i (t)} + \mathbf {E}^{*} (\omega, t) \left. \frac{d \chi_{\mathbf Q}(\omega) }{d \mathbf Q}\right|_{\mathbf Q= \mathbf Q_i(t)}  \mathbf{E}(\omega, t),
    \label{eq:Motion}
\end{align}  
\end{widetext}

where $U$ is the total energy without illumination, $\mathbf{E}(\omega, t) = \mathbf{A}(t)e^{i\omega t}\simeq \mathbf{A}(t) $ is the electric field of the light approximated by its slowly varying (when compared to $1/\omega$) amplitude $\mathbf{A}(t)$, and $\chi_{\mathbf Q}(\omega)$ is the macroscopic polarizability tensor of the system at coordinates ${\mathbf Q}$. Both $U$ and $\chi_{\mathbf Q}$ are obtained from DFT and TDDFT calculations on a dense grid of configurations, and interpolated using a radial basis interpolation on that grid following the procedure detailed in Supplementary information.

\begin{figure}[h!]
    \centering
    \includegraphics[width= 0.95\linewidth]{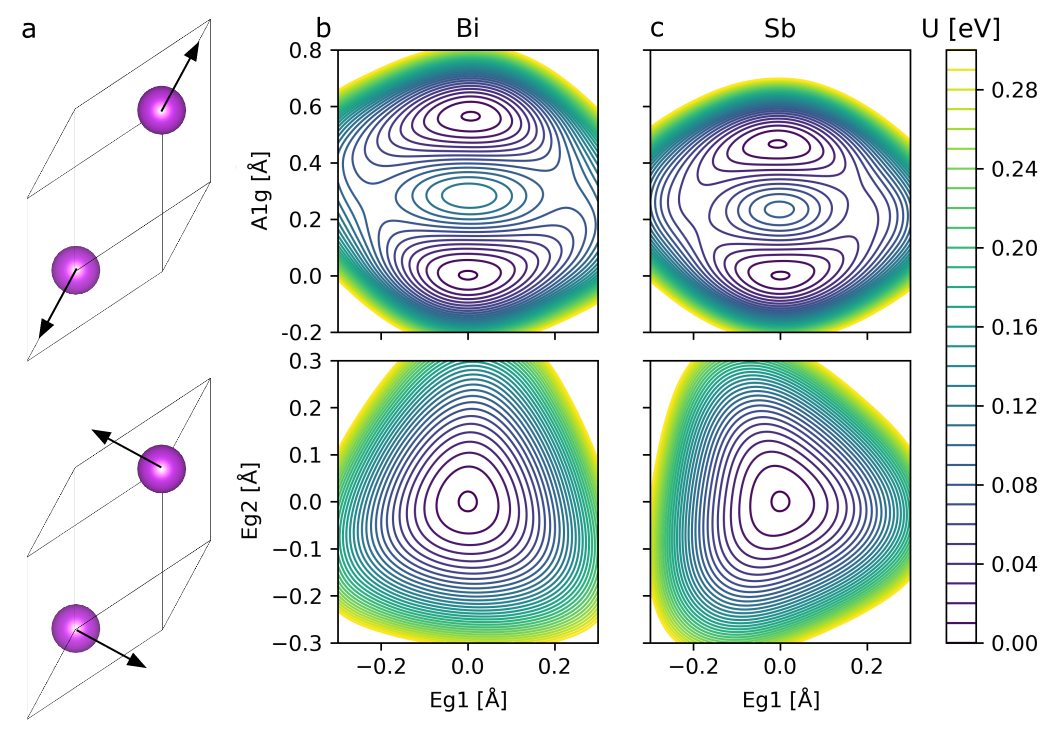}
    \caption{a) The orthorhombic unit cell of Bi and Sb with the motion of atoms associated with the $A_{1g}$ (top) and one of two degenerate $E_{g}$ phonon modes (bottom). In panels b and c, we show the potential energy surface with atoms moved along pairs of normal modes, with the third normal mode set to $Q=0$. Coordinates (0,0) correspond to the ground-state configuration. Potential energy (U) is in units of  eV/unit cell (2 atoms). }
    \label{fig:f1}
\end{figure}

%\section{Results and Discussion}
Both Bi and Sb form crystals with an orthorhombic unit cell consisting of two atoms with a three-fold symmetry around the body diagonal (see Figure~\ref{fig:f1}). Thus, there are two point group symmetries associated with the three optical phonon modes of interest in this work: the fully symmetric $A_{1g}$ mode, corresponding to an out-of-phase displacement of the two atoms along their bond, and two doubly-degenerate $E_g$ modes, corresponding to displacements normal to that bond. As shown in Ref. \citenum{merlinSSCGeneratingCoherentTHz1997}, only the Raman tensor of $A_{1g}$ is diagonal (symmetric under all orthorhombic point group operations), in contrast with the off-diagonal Raman tensor of the $E_g$ modes, suggesting the latter can only be excited via second-order effects. 

 Figure~\ref{fig:f1} presents the PES as a function of atomic motion along the $A_{1g}$ and  $E_g$ optical zone-centered phonon modes of Bi and Sb. Both materials have a characteristic double-well potential along their $A_{1g}$ direction with both minima being structurally equivalent. This energy profile is a consequence of the broken symmetry ground-state due to a Peierls-like distortion from a high symmetry configuration (at $Q_{A1g} \simeq 0.3$ \AA), where no Bi-Bi and Sb-Sb bond is elongated. The predicted dimerization energy (energy difference between a minimum and the high-symmetry configuration per 2-atom  unit cell) for Sb ($\sim$ 100 meV) is lower than that for Bi ($\sim$ 140 meV), indicating that Bi has a stronger tendency towards dimerization, and, accordingly, that the $A_{1g}$ mode in Sb near the energy minima are more anharmonic. In contrast, $E_g$ modes retain a near-harmonic character for displacements $Q_{Eg} \simeq \pm 0.2$Å  with the trigonal deviation caused by the three-fold symmetry of the lattice. In the limit of small displacements, we predict the phonon frequencies for the $A_{1g}$ and  $E_g$ modes to be, respectively (3.06, 2.47) THz for Bi and (4.31, 2.77) THz for Sb, in good agreement with the experimental values of (2.9, 2.2) (4.5, 3.5) THz~\cite{chengAPLImpulsiveExcitationCoherent1990}.

A simple quantification of the mode anharmonicity can be obtained by fitting the energy vs. displacement curve for a single phonon coordinate (the two others being fixed to $Q=0$, neglecting intermode coupling) by a higher-order polynomial: 
For $A_{1g}$, $E_{g1}$, and $E_{g2}$, respectively, we obtain the following third-order and fourth-order coefficients for Bi  and Sb are shown in Table~\ref{tab:Anharmonic_coefficients}.
%(-4.307(21) eV/\AA$^3$,  0.580(7) eV/\AA$^3$, -2.120(11) eV/\AA$^3$; -8.6(4) eV/\AA$^4$, 2.78(7)eV/\AA$^4$, 1.9(1)eV/\AA$^4$)
%and for Sb (8.555(22) A1geV/\AA$^3$,  2.192(25) eV/\AA$^3$, 0.908(13) eV/\AA$^3$; -21.6(4) A1g eV/\AA$^4$, 7.05(24)eV/\AA$^4$, 8.03(13)eV/\AA$^4$). 
The magnitude of these coefficients is associated with two main effects: First, as the force depends on the distance from the equilibrium position, the phonon frequency becomes amplitude-dependent, an effect observed experimentally for the $A_{1g}$ mode~\cite{decampPRBDynamicsCoherentControl2001} in Bi. Second, as the third and fourth order force constants correspond to 3- and 4-phonon interactions, these coefficients are related to  the onset of higher harmonics at integer multiples of the original frequencies. Beyond these two effects, our approach also contains cross-phonon terms (e.g. $Q_{A_{1g}} Q_{E_{g}}^2$) that transfer energy across the different branches, which enable selective excitation of originally-degenerate  phonon modes (though observing this effect can be complicated by the large decay rates of these modes \cite{haseAPLOpticalControlCoherent1996}).

\begin{table}[]
    \centering
    \begin{tabular}{|c|cc|cc|}
\hline
        & \multicolumn{2}{c|}{Bi} & \multicolumn{2}{c|}{Sb}\\
        & $Q^3$ [eV/\AA$^3$] & $Q^4$ [eV/\AA$^4$] & $Q^3$ [eV/\AA$^3$] & $Q^4$ [eV/\AA$^4$]\\
\hline
      $A_{1g}$ & -4.3 & -8.6 &8.6& -21.6\\
      $E_{g1}$ & 0.6 & 2.8 & 2.2 & 7.1\\
      $E_{g2}$ & -2.1 & 1.9 & 0.9 & 8.0\\
     \hline 
    \end{tabular}
    \caption{Third (Q$^3$) and fourth (Q$^4$) order fit coefficients of the potential energy surface with displacements along the three phonon normal modes for both Bi and Sb.}
    \label{tab:Anharmonic_coefficients}
\end{table}

\begin{figure*}[htb!]
    \centering
    \includegraphics[width= \textwidth]{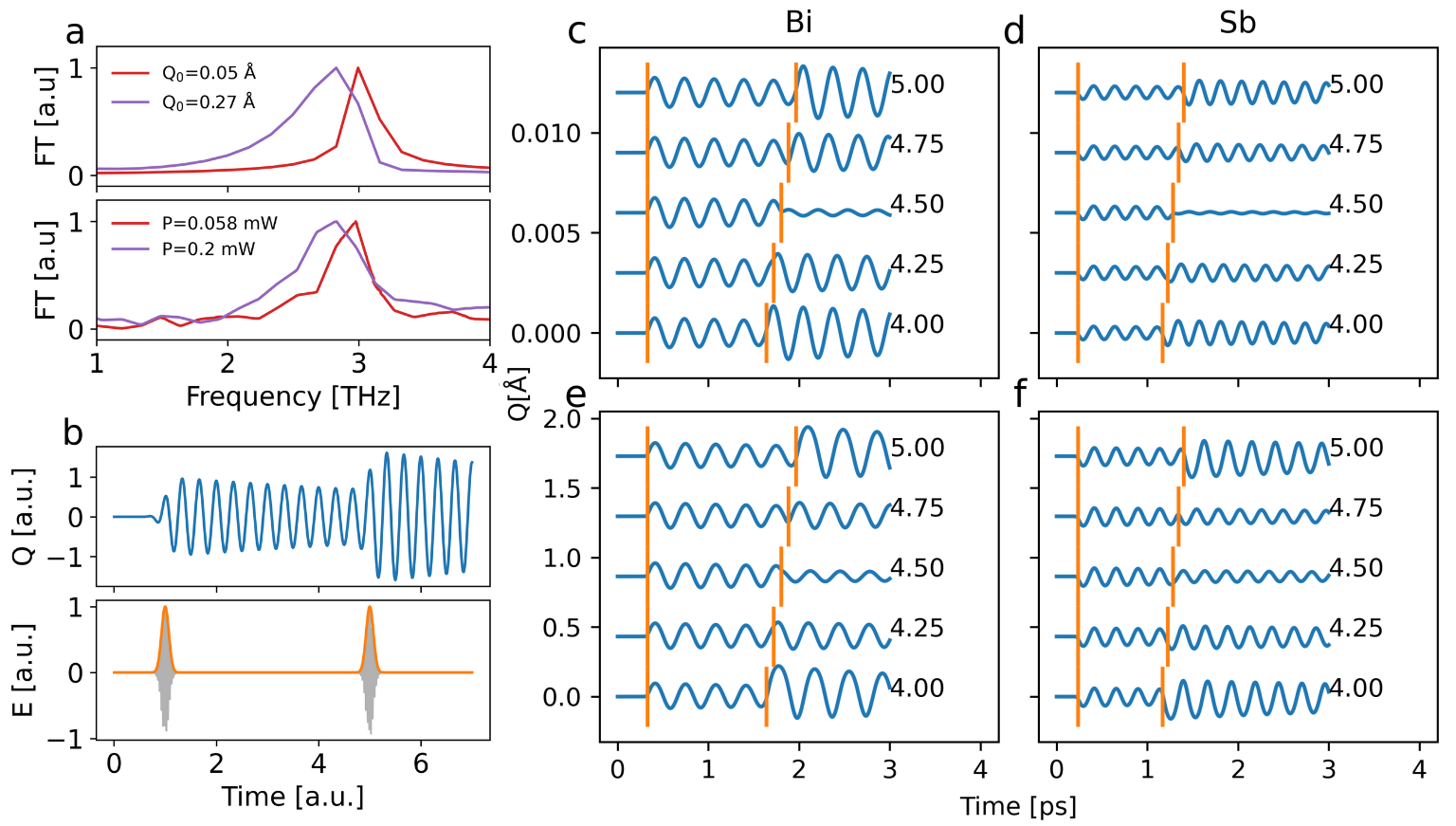}
    \caption{a) Comparison of experiment (bottom) and theory (top) in the extracted frequency of the coherent phonon within Bi via the Fourier transform of the reflectivity trace at two different laser fluences/simulated amplitude. The theoretical displacement is chosen to reproduce the frequency shift observed in experiment. b) Schematic of a 2-pulse experiment and a resulting oscillatory trace. c-f) show the 2-pulse dynamics with a second pulse placed half-integer time period away in order to fully destructively interfere with the first pulse within the harmonic approximation. c) and d) show waterfall plots of the trajectories simulated for low pulse powers in Bi and Sb. $Q$ is the amplitude along the $A_{1g}$ coordinate, and the plots are shifted by a different constant for the different trajectories. Different trajectories are labeled by the time-delay in units of the harmonic phonon time period (4.0 T - 5.0 T). e) and f) show the same trajectories generated by a high pulse powers as defined in the text. A phenomenological damping rate of 0.1 THz was added to the simulations.}
    \label{fig:f2}
\end{figure*}

To analyze the effect of the mode anharmonicity on the light-induced dynamics, we conduct transient spectroscopy experiments on bismuth at various pump laser intensities. All experimental details are indicated in supplementary information. We then extract the frequency domain response from the pump-probe traces, showing the pump-intensity-dependent shift of the frequency of the $A_{1g}$ phonon response. At low fluence, the coherent phonon frequency is in good agreement \cite{chengAPLImpulsiveExcitationCoherent1990} with the one measured through Raman spectroscopy. However, and as shown in Fig.~\ref{fig:f2} a), higher laser fluences lead to a softening of the $A_{1g}$ phonon frequency, in agreement with prior measurements~\cite{decampPRBDynamicsCoherentControl2001,hasePRLDynamicsCoherentAnharmonic2002,lanninPRBSecondorderRamanScattering1975}. 
As discussed above, such behavior can be explained by the larger displacement at higher fluences, giving rise to larger anharmonicity. We also note the appearance of second harmonics for a laser power of 0.2 mW (see supplementary information), while the $E_{g}$ modes are predicted and observed to be negligibly excited for non-polarized light. We note that for our system, the phonon frequencies are well separated and so we do not expect multiple peaks in the Reflectivity spectrum.~\cite{Sun_2017}

As we will now show, the phonon softening has important consequences for phonon excitation protocols. We illustrate this point by simulating the light-induced dynamics caused by a simple, 2-pulse protocol, shown in Figure~\ref{fig:f2} b), where the two pulses are separated by a delay expressed as a multiples of the phonon time period. For purely harmonic potentials, pulses sent at integer phonon time periods lead to constructive interference between the two coherent excitations, i.e. phonon amplification, while pulses sent at half-integer periods lead to destructive interference, i.e. phonon annihilation~\cite{haseAPLOpticalControlCoherent1996,decampPRBDynamicsCoherentControl2001,dekorsyLSiSVCoherentPhononsCondensed2000}. Accordingly, and as shown in Fig.~\ref{fig:f2} c) and d), excitations at weak electric field amplitude (4.2 MV/m and 2.1 MV/m for Bi and Sb respectively, corresponding to a peak laser power of  2.9 kW and 1.5 kW respectively, see supplementary information) result in low-amplitude phonon oscillations, and such phonons can be amplified and annihilated at times based on their integer/half-integer oscillation periods. 

However, at the stronger electric field amplitude of $\sim$ 67 MV/m and $\sim$ 54 MV/m for Bi and Sb respectively, (corresponding to a peak laser power of  748 kW and 486 kW respectively), the amplitude of the oscillation ($\sim$0.1-0.2 \AA) leads to significant anharmonic effects: As shown in Fig.~\ref{fig:f2} e) and f), the corresponding oscillation period becomes longer, with 4 full oscillations taking $\sim$4.1 harmonic oscillation periods. Considering the frequency, $\omega$, at any intensity to take the form,

\begin{align}
    \omega = \omega_0 + \delta \omega (A(t)),
\label{Eq:3}
\end{align} 

where $\omega_0$ is the low-amplitude phonon frequency, respectively, and $\delta \omega$ is the change in frequency due to anharmonicity, with $\delta \omega (A(t)) \simeq 0.2$ THz at $\simeq 0.2$ \AA. 

We note that an analytical solution to Eq.~\ref{Eq:3} is possible provided a known analytical form of the PES. For example, Ref.\cite{iyikanatANNonlinearTunableVibrational2021} introduced an expression for $\delta \omega (A(t))$ for the case of hexagonal boron nitride, where the potential energy surface along the infrared-active phonon coordinates can be accurately described by a fourth order polynomial. 

\begin{figure}[h!]
    \centering
    \includegraphics[width=0.95 \linewidth]{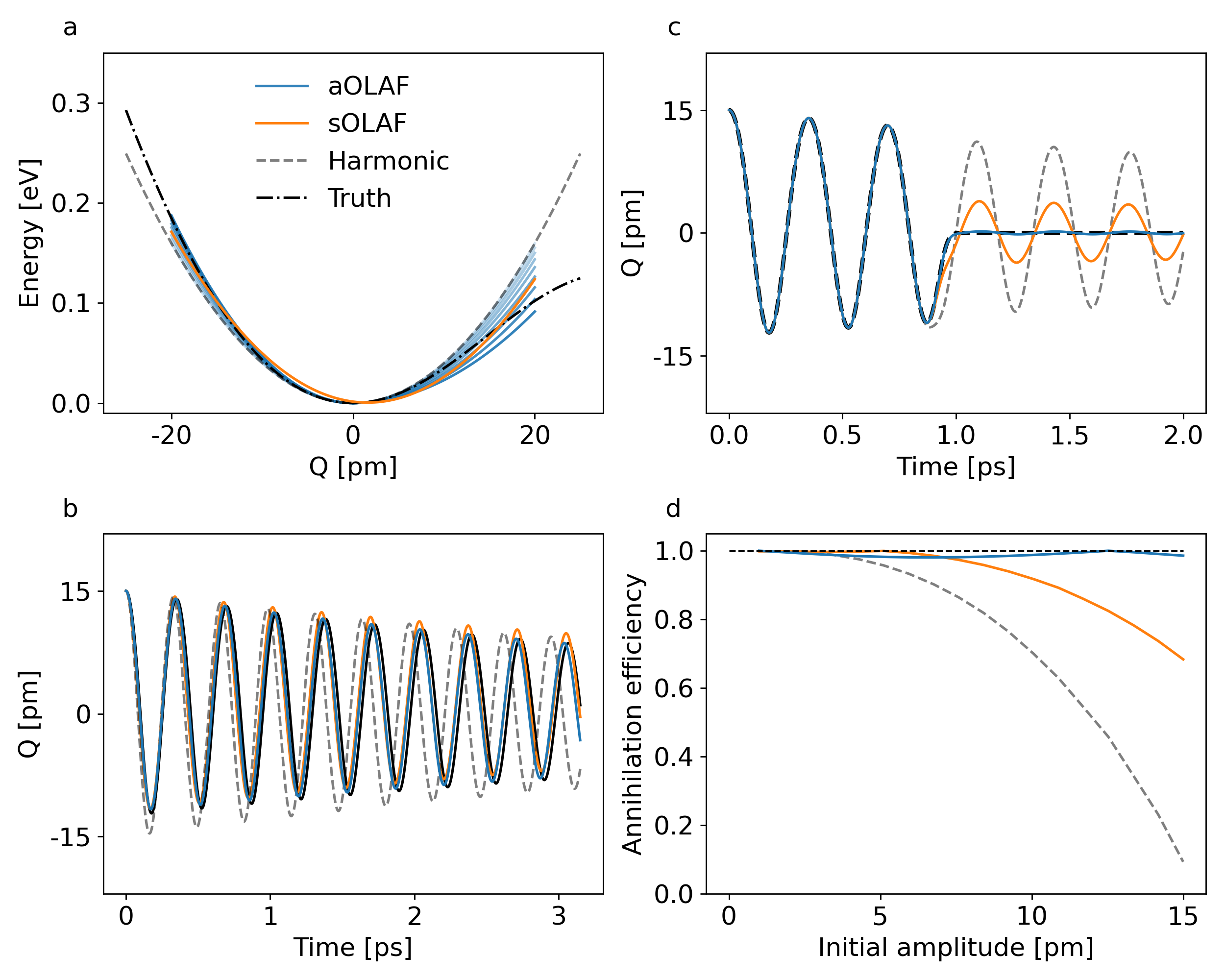}
    \caption{a) The calculated PES along the $A_{1g}$ coordinate compared with the harmonic approximation of the PES, the symmetric (s), and asymmetric (a) effective harmonic approximation (OLAF) introduced in the text. For aOLAF, we plot nine lines with differing amplitudes uniformly spaced by 2.5 pm and ranging from 0 pm to 20 pm, shaded such that the darker the colors correspond to larger amplitudes. b) Phonon trajectories initialized with an amplitude of 10 pm, corresponding to the ground truth PES, harmonic, and two OLAF approximations. c) Coherent phonon annihilation via 2-pulse excitation within the three approximations for an initial amplitude of 15 pm. d) The proportion of the oscillation amplitude that is removed after the second pulse.}
    \label{fig:f3}
\end{figure}

Instead, we now propose a simple correction to standard protocols that enable their extension to the anharmonic regime. We use an approximation inspired by the amplitude-dependent harmonic approximation as described in Ref. \cite{denmanAJoPAmplitudeDependenceFrequencyLinear1959}. Here, we approximate the PES by a harmonic potential \textit{dependent on the amplitude}, i.e. $E({\mathbf Q}) =  1/2 \left.\frac{dU}{d \mathbf Q}\right|_{\mathbf Q} \left| Q \right|^2  $. Under this assumption, the amplitude-dependent force constant is an instantaneously linear approximation to the restoring forces that the oscillator feels over its trajectory. We refer to this protocol as the optimal linear approximation of forces (OLAF). Importantly, by enabling $\left.\frac{dU}{d \mathbf Q}\right|_{\mathbf Q}  \neq \left.\frac{dU}{d \mathbf Q}\right|_{- {\mathbf Q} }$, OLAF can in principle reproduce the large and amplitude-dependent asymmetry of the PES along the $A_{1g}$ coordinate. Along a given coordinate, we obtain the amplitude-dependent  restoring force by minimization of
\begin{align}
\mathbf F_{OLAF} (A) =  min(||\mathbf{F} -  \mathbf{F}_{truth} (|Q| < A) ||_2), \label{Eq:FunctionalMin}
\end{align}
 for $Q$ in $[0,+A]$ at positive $Q$ and $Q$ in $[-A,0]$ at negative $Q $, with $A$ being the instantaneous amplitude. Importantly and as shown in Supplementary Information (Fig S2), these values can also be approximated by independently fitting the half-period at positive and negative amplitude. Eq. \ref{Eq:FunctionalMin} can be understood as an approximating the full potential energy surface defined by $\mathbf{F}_{truth}$ by a series of amplitude-dependent harmonic potentials $\mathbf F_{OLAF} (A)$ extracted from each half-period of oscillation. 

We note that this approach has key differences when compared with temperature dependent effective potentials (TEDP). TDEP studies have generally focused on the study of acoustic modes, which have a different behavior than optical modes that we study here. Additionally, it is important to note that while we neglect thermal expansion, we expand the potential energy surface at higher orders than the harmonic approximation, which TDEP approximates.

%Moreover, in contrast to Ref. \citenum{denmanAJoPAmplitudeDependenceFrequencyLinear1959},   where a Chebyshev approximation of the nonlinear potential is used, we use a least-squares approximation to the gradient of the PES, fitted to a line \cite{hellmanPRBLatticeDynamicsAnharmonic2011}. 

Figure \ref{fig:f3} a) presents the difference between the harmonic approximation and the OLAF approach along the $A_{1g}$ coordinate, with the amplitude varied in $\pm$ regime as was found to agree well with experiment. For OLAF, we plot the second-order polynomial fits to the PES for Q within $\pm$20 pm, and contrast it to the ground truth, an harmonic fit obtained analytically as a second derivative at Q = 0, and the OLAF protocol under the constraint of $\left.\frac{dU}{d \mathbf Q}\right|_{\mathbf Q}  = \left.\frac{dU}{d \mathbf Q}\right|_{- {\mathbf Q} }$ (labeled as symmetric OLAF). As expected, the harmonic approximation deviates from the ground truth as the amplitude approaches the high-symmetry configuration. While a symmetrized amplitude-dependent effective harmonic potential can extend the domain of validity of the harmonic approximation, this is accomplished by shifting the energy minimum away from equilibrium towards more positive amplitudes for larger displacement. In contrast, OLAF conserves the correct position of the energy minimum and accounts for the asymmetry between positive and negative amplitudes. We note, however, that OLAF can only indirectly reproduce the concavity of the PES at positive amplitudes through a set of rapidly decaying force constants.     

Importantly, our amplitude-dependent harmonic approximation leads to a trajectory with smaller deviation from the true trajectory over a longer period of time as shown in Fig. \ref{fig:f3} b). We note that the softening of the phonon mode can be seen through the change in oscillation time period in the truth as well as the OLAF trajectory, but is missing from the harmonic approximation. Another important consequence of the energy profile along the $A_{1g}$ coordinate is the shorter half-period for the negative amplitudes as compared to the positive amplitudes, as well as their inverse behaviors at large amplitudes: As shown in Fig S2, the negative-amplitude half-periods are slightly shortened at high amplitude, while the positive-amplitude half-periods are increased. This phenomenon is captured quantitatively by the asymmetric fit, and can also be derived by measuring the lengths of the half-periods through the zero crossings of the experimental spectroscopic traces.  

The consequences of these approximations in determining two-pulse protocols are shown in Figure~\ref{fig:f3}c, for a standard phonon-annihilation protocol shining a second pulse 2.75 periods after an amplitude maximum efficient at small amplitude: Over the course of those 15 pm oscillations, the phase accumulated in the harmonic approximation considerably reduces the effectiveness of the protocol. In contrast, the 2.75 periods predicted by OLAF closely match the numerically-optimized result. We note that the asymmetry between the positive and negative amplitudes is enough to significantly lower the efficiency of the symmetrized OLAF protocol at those amplitudes, indicating that our asymmetric protocol is particularly suited to broken-symmetry materials.

We define the figure of merit corresponding to the destructive interference as:
\begin{align}
    \mathcal{F} = 1 - \frac{A_{2p}}{A_0}
\end{align}

where $A_{2p}$ is the time-averaged norm of the oscillations after the second pulse and $A_0$ is the same quantity without the second pulse.
In the case of the full PES, the annihilation efficiency would be 1. However, for the harmonic approximation the efficiency decays from 0.9 at 5 pm to $<0.7$ at 10 pm. The OLAF effective harmonic approximation improves the validity, extending the range of annihilation up to 10 pm. In contrast, the asymmetric OLAF protocol enables full annihilation efficiency across this range of amplitudes.

Modeling oscillations with amplitude-dependent frequencies and the approximate generalization to damped oscillations provides a compromise between fully anharmonic dynamics which have no closed form, and harmonics dynamics using Taylor-expanded potential energies, which have a closed form but a limited range of accurate validity: as shown in Figure S1, the PES computed in Figure \ref{fig:f1} and fourth order polynomial fit of the PES proposed by Ref. \citenum{iyikanatANNonlinearTunableVibrational2021} result in accurate frequency shifts as a function of amplitude for large amplitudes (see SI, Section III). In contrast, aOLAF deviates from the exact solution by only $0.02$ THz at a 10pm amplitude yet by $\sim 0.15$ THz at a 20 pm amplitude.  In general, the parametrization of the aOLAF protocol through the half-periods of the oscillations will be efficient as long as there are no more than one zero-derivative point in the PES crossed over the course of the trajectory, i.e. as long as the energy of the oscillator is smaller than the energy barrier.

\section{Conclusion}
In conclusion, we have quantified the role of anharmonic effects in the lattice response of bismuth and antimony, and proposed a protocol for controlling them. We developed a DFT-informed model of the light-induced lattice dynamics, using an interpolated potential energy surface along the three optical phonon modes. With this model, we investigated the low- and high-amplitude frequencies of light-induced phonons in these materials using a field-dependent, Born-Oppenheimer description of the potential energy surface. We  introduced a framework that accurately captures the amplitude-dependent frequency and accumulated phase of the coherent phonon via linearization of the amplitude-dependent frequency of the coherent phonon. The formalism and results presented provide a way to accelerate the calibration of two-pulse experiments and improve the description of protocols for nonlinear phononics.

%Supporting Information. Details of implementation of the classical equation of motion, the measured Raman spectrum for Bi, and the momentum and frequency resolved dielectric function.

\section{Acknowledgement}
SS and AH acknowledge financial support from the U.S. Department of Energy (DOE), Office of Science, Basic Energy Sciences under Award No. DE-SC0023402. PD, AH and ZH acknowledge support from the U.S. Department of Energy, Office of Science, BES Microelectronics Threadwork, under contract number DE-AC02-06CH11357. Use of the Center for Nanoscale Materials, an Office of Science user facility, was supported by the U.S. Department of Energy (DOE), Office of Science, Office of Basic Energy Sciences, under Contract DE-AC02-06CH11357.

\bibliography{refs}
\end{document}